\documentclass[a4paper]{article}

\usepackage{INTERSPEECH2022}
\usepackage{amsmath,graphicx}
\usepackage{hyperref}
\usepackage{url}
\usepackage{caption}
\usepackage{subcaption}
\usepackage{microtype}
\usepackage{booktabs}
\usepackage{tabularx}
\usepackage{multirow}

\newcolumntype{C}{>{\centering\arraybackslash}X}
\newcolumntype{L}{>{\raggedright\arraybackslash}X}
\newcolumntype{R}{>{\raggedleft\arraybackslash}X}
\newcolumntype{P}[1]{>{\raggedright\arraybackslash}p{#1}}
\usepackage{siunitx}
\usepackage{etoolbox}   
\newcommand{\ubold}{\fontseries{b}\selectfont}  
\robustify\ubold     

\newcommand{\tablecaptionsep}{\vspace*{-5pt}}

\usepackage{xcolor}
\definecolor{hermancolor}{HTML}{FF6600}

\definecolor{mattcolor}{HTML}{0004ff}

\definecolor{shorten}{HTML}{7a7a7a}

\title{Voice Conversion Can Improve ASR in Very Low-Resource Settings}
\name{Matthew Baas and Herman Kamper}
\address{MediaLab, E\&E Engineering, Stellenbosch University, South Africa}
\email{20786379@sun.ac.za, kamperh@sun.ac.za}

\begin{document}
\ninept
\maketitle
\begin{abstract}
Voice conversion (VC) could be used 
to improve speech recognition systems in low-resource languages by using it to augment limited training data.
However, VC has not been widely used for this purpose because of
practical issues such as 
compute speed and limitations when converting to and from unseen speakers.
Moreover, it is still unclear whether a VC model trained on one well-resourced language can be applied to speech from another low-resource language for the aim of data
augmentation.
In this work we assess whether a VC
system can be used cross-lingually
to improve low-resource speech recognition.
We combine several recent techniques to design and train a practical VC system in English, 
and then use this system to augment data for training speech recognition models in several low-resource languages.
When using a sensible amount of VC augmented data, 
speech recognition performance is improved in all four low-resource languages considered.
We also show that VC-based augmentation is superior to SpecAugment (a widely used signal processing augmentation method) in the low-resource languages considered.

\end{abstract}
\noindent\textbf{Index Terms}: voice conversion, data augmentation, 
low-resource speech processing, speech recognition.

\section{Introduction} \label{sec:intro}

Although
automatic speech recognition (ASR) systems
have 
greatly improved over the last few years~\cite{wav2vec2.0, asrsota-zhang2020pushing, hubert2021},
they still struggle in very low-resource settings
where
training data is minimal and a high quality language model (LM) is not available~\cite{besacier2014asr_survey}.
Concretely, even state-of-the-art ASR performance is limited for
languages where we only have a few minutes of transcribed audio from a handful of speakers~\cite{xlsr-wav2vec2}, e.g.\ when building systems for regional dialects or code-switched speech.

One possible solution is to use data augmentation: available data is processed so that it has different characteristics from the original while preserving the content (words spoken), effectively giving more training data.
This idea has a long history~\cite{Hunt1989data_aug,Lathoud2005data_aug}, with more recent methods like SpecAugment~\cite{park2019specaugment} improving ASR performance and robustness.
While these signal processing approaches have proven effective, they are mostly constrained to direct general-purpose modifications of the audio waveform --\ e.g.\ pitch shifting, time stretching and frequency masking\ --\ ignoring some of the unique aspects of human speech.
To address this, one research direction is pursuing the idea of using voice conversion for data augmentation.

In voice conversion (VC), an utterance spoken by one speaker is processed so that it sounds as if it was spoken by a different target speaker, while keeping the content unchanged~\cite{Mohammadi2017vc_def}.
VC approaches are becoming more practical in terms of speed~\cite{hifi-gan}, naturalness~\cite{vcc2020, triple_bottle_discrete_disentangle}, and performance on unseen speakers~\cite{autovc, stargan-zsvc}.
Some newer methods even provide finer-grained
ways to control different properties of the generated speech~\cite{hst,triple_bottle_discrete_disentangle}, and can be used cross-lingually~\cite{yang2021cross}.

To use VC for data augmentation,
ASR training utterances are fed into a VC system and converted to arbitrary target speakers, thereby creating additional synthetic training utterances with a known transcript.
It has been hypothesized that this would improve ASR performance in low-resource settings~\cite{jaitly2013vltp_aug}
, although initial attempts (Sec.~\ref{sec:related}) have had 
limited success.
Moreover, no
studies have considered applying a VC model trained on a well-resourced language to speech from another unseen low-resource language for the purpose of data augmentation --\ this is a requirement for using VC augmentation in arbitrary settings.

Here we
aim to address this shortfall and see if VC can indeed be used as an effective data augmentation approach to improve ASR in unseen low-resource settings. 
We first propose a VC system trained on English that is practical for cross-lingual data augmentation: running faster than real time on unseen speakers and unseen languages.
We then apply this VC approach to create synthetic data for several low-resource settings in English, Afrikaans, Sepedi, isiXhosa and Setswana.
To answer our research question, we compare the
performance of
ASR systems trained with and without generated voice-converted data, and also compare to SpecAugment.
With limited training data (10 min) we find that our VC approach is complementary to SpecAugment on English, while it is superior for data augmentation on the low-resource languages.
To the best of our knowledge, we are the first to use cross-lingual VC for data augmentation, where VC is applied to unseen languages and speakers.

\section{Related work}  \label{sec:related}
Laptev et al.~\cite{tts_data_aug} explored using fast text-to-speech (TTS) methods to perform data augmentation, obtaining 
improvements 
on medium-resource ASR.
Since terms about resources are contested in the literature~\cite{how_low_is_low_resource}, we define
\textit{very low-resource}, \textit{low-resource}, and \textit{medium resource} as settings with roughly 10~min, 1~h, and 10~h to 50~h of labelled audio, respectively.
However, for very low-resource applications like
the ones explored in this paper, a good TTS model or LM is usually unavailable.

The idea of using VC to augment data has a long history~\cite{Bellegarda1994spk_norm_vc}.
Cui et al.~\cite{2015_simple_vc_for_asr} made initial attempts to improve {medium-resource} ASR 
showing moderate improvements despite limitations of VC techniques available at the time.
Other studies~\cite{2021vc_for_tts, 2020children_vc_asr} showed that using more recent deep learning-based VC methods can improve TTS systems where the desired speaker is low-resource.
However, these studies use a target speaker from a language seen during training, leaving the question open as to whether ASR can be improved in unseen low-resource languages. 

Zhao et al.~\cite{crosslingual_vc} made an attempt to use cross-lingual VC to improve code-switched TTS models. 
However, their VC models are trained separately 
for each speaker considered, making their system unsuitable for use on new unseen low-resource languages.
Finally, \cite{2019failed_vc_for_asr} found that applying a deep-learning-based VC approach failed to noticeably improve 
medium-resource
ASR models when all models are trained and evaluated on English. 
They were, however, limited by practical problems associated with VC approaches
available at the time, {such as the fast but low-quality Griffin-Lim vocoder}.

\section{Voice conversion method} \label{sec:format}
Our goal is to use VC for low-resource data augmentation.
This means that our overall VC model needs to be computationally efficient, and be able to convert input speech from a language (and speaker) that are unseen during training.

\begin{figure}[!t]
    \centering
    \hspace{0.05\linewidth}
    \begin{subfigure}[b]{0.94\linewidth}
      \centering
      \centerline{\includegraphics[width=0.75\textwidth]{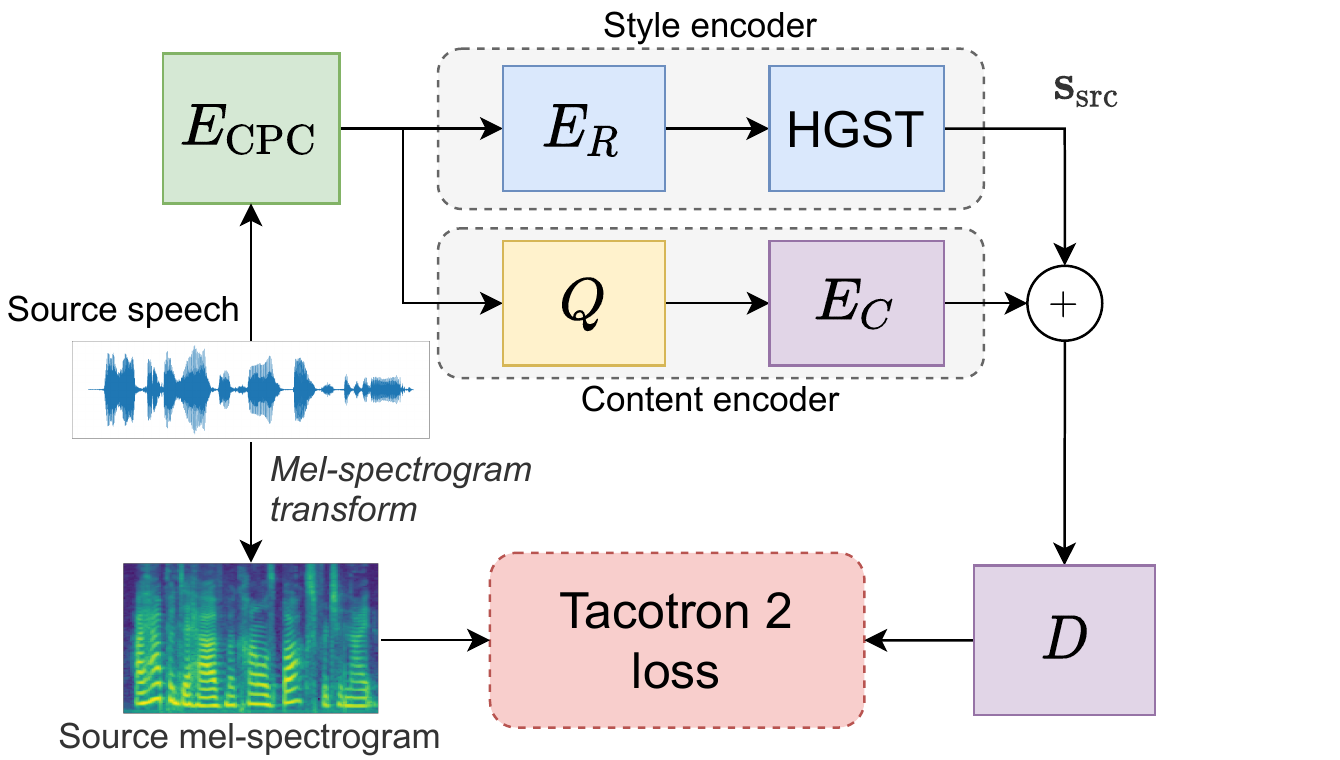}}
      \caption{Training setup}\label{fig:block_diag_train}\medskip
    \end{subfigure}
    \vspace*{-2.5pt}
    \hspace{0.05\linewidth}
    \begin{subfigure}[b]{0.94\linewidth}
      \centering
      \centerline{\includegraphics[width=0.69\textwidth]{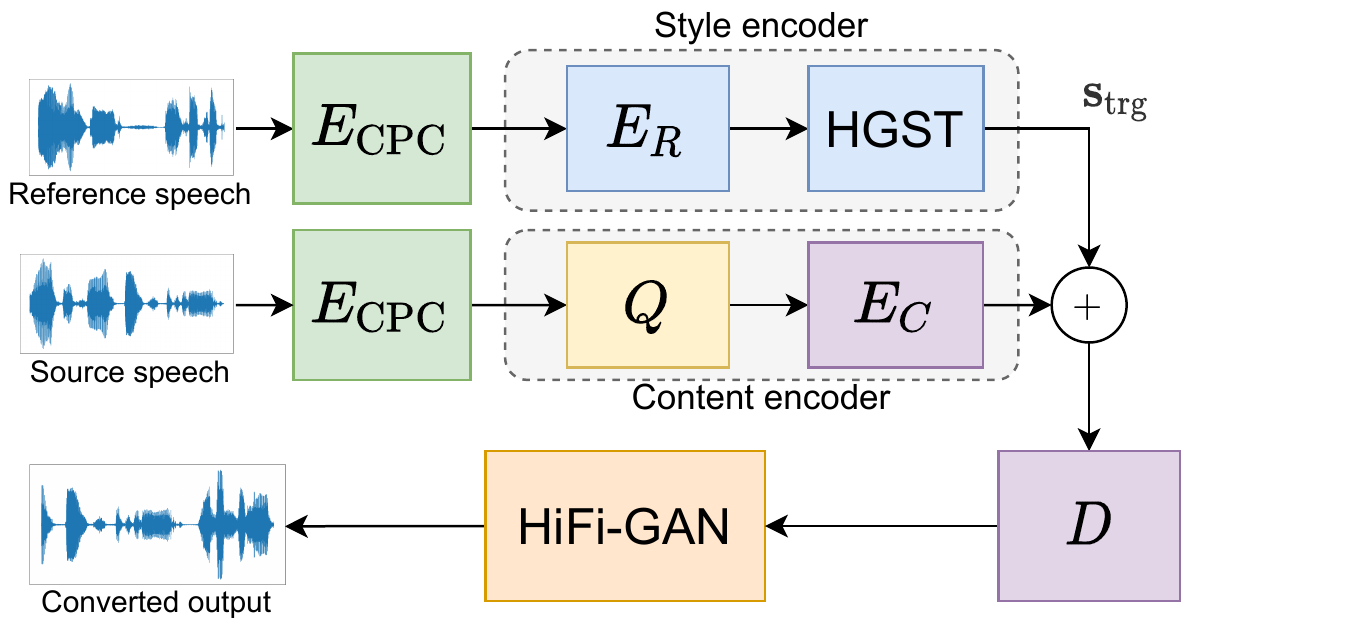}}
      \caption{Inference setup}\label{fig:block_diag_infer}
    \end{subfigure}
\vspace*{-10pt}
\caption{
Our VC approach.
(a)~During
training, the same utterance is fed through a CPC encoder into both the style and content encoders; the decoder tries to
reconstruct the input utterance's spectrogram. 
(b)~At
inference, separate input and reference utterances are fed into the style and content encoders, respectively, before vocoding the spectrogram using HiFi-GAN~\cite{hifi-gan}.}
\label{fig:block_diag}
\vspace*{-7.5pt}
\end{figure}

\textbf{Overview:} 
Our approach is shown in Fig.~\ref{fig:block_diag}.
We want to convert an input waveform into a sequence of feature vectors and then separate the \textit{linguistic content} (the words being spoken) from the \textit{speaker information}.
We do this with specially designed style and content encoders, denoted with  dashed boxes.
The style encoder produces a single \textit{style vector} $\mathbf{s}$, while the content encoder produces a sequence of \textit{content embedding vectors}.
During training (Fig.~\ref{fig:block_diag_train}) the style vector $\mathbf{s} = \mathbf{s}_\textrm{src}$ and content embeddings are obtained from the same input utterance. The model is trained to reconstruct the input by summing $\mathbf{s}_\textrm{src}$ with each content embedding, passing the resulting vectors into a decoder module.
At test time (Fig.~\ref{fig:block_diag_infer}) the content and speaker encoders receive speech from different utterances.
This means that the source or reference utterance (or both) may be from languages and speakers unseen during training.
If the content and speaker information is appropriately separated, then the source utterance will be produced in the voice of the speaker supplying the reference utterance. Details on each component~follows.

\textbf{Speech encoder:}
\mbox{Instead of using a spectrogram represent} the reference or source utterances, we use a pretrained speech encoder.
Specifically, we use the contrastive predictive coding (CPC) model from~\cite{nguyen2020zero}.
This model, $E_{\textrm{CPC}}$ in Fig.~\ref{fig:block_diag}, produces a 512-dimensional vector for every 10~ms of 16~kHz-sampled audio.
This model transforms the input waveform into representations that makes it easier to disentangle content and speaker information for the downstream style and content encoders~\cite{niekerk21_interspeech}.

\textbf{Style encoder:}
The style encoder consists of a reference encoder $E_R$ and a
hierarchical global style token (HGST) layer~\cite{hst}.
The reference encoder $E_R$ consists of several 1-D convolutional layers followed by an LSTM \cite{lstm}.
The final hidden state of the LSTM is linearly transformed and passed to the HGST layer.
The HGST layer acts as a form of information bottleneck;
the idea is that the network learns to \textit{only} include speaker information in $\mathbf{s}$, since the content information can be obtained from the content encoder output.
The HGST layer is parameterized by $l$ sublayers and $h$ trainable vectors per sublayer.
The first sublayer attempts to represent the last output vector from $E_R$ as a convex combination of its $h$ vectors.
The difference between this combination and the input vector is then used as the input to the next sublayer, so each layer in the hierarchy
models the residue from the previous approximation~\cite{hst}.
With small $h$ and $l$, the HGST layer forces the final style vector $\mathbf{s}$ to only retain information about the global speaking style. 

\textbf{Content encoder:}
The content encoder attempts to remove speaker information from the CPC features so as to only retain linguistic content.
The quantization block $Q$ in Fig.~\ref{fig:block_diag} first performs speaker normalization using means and variances computed for features from the pretrained $E_{\textrm{CPC}}$ on a subset of the training data.
This was shown to improve speaker invariance~\cite{niekerk21_interspeech}.
During inference with unseen source speakers, we compute the mean and variance on the features of the single source utterance.
The normalized embeddings are quantized by representing them by the nearest centroid vector in a $K$-means clustering model with $K=100$.
This quantized sequence is then passed through a content encoder module $E_C$ with the same architecture as $E_R$, except returning all LSTM states -- one content vector every 10~ms.
The $K$-means and speaker normalization parameters are precomputed before the rest of the network is trained.

\textbf{Decoding and vocoding:}
Each content vector is summed with the style vector $\mathbf{s}$.
The resulting sequence is then passed into a Tacotron~2 decoder network $D$ to produce an output mel-spectrogram~\cite{tacotron2}.
We use the same Tacotron~2 decoder architecture as in~\cite{tacotron2}.
During training (Fig.~\ref{fig:block_diag_train}), the loss is formulated the same as for the original Tacotron 2: an $L_2$ loss between the predicted spectrogram and the spectrogram of the source utterance
with an additional
loss associated with the length of the output spectrogram~\cite{tacotron2}.
During inference (Fig.~\ref{fig:block_diag_infer}), the output spectrogram is converted to the time-domain using
We use the {spectrogram parameters} and pretrained HiFi-GAN model from~\cite{hifi-gan}, due to its high performance and fast inference speed.

With these choices for the different modules, our overall VC approach is faster than real time on a modern GPU.

\section{Experimental setup}\label{sec:exp_setup}

\textbf{Data:}
We want to know whether a VC model trained on a well-resourced language can be applied to an unseen low-resource language to improve ASR.
English is our well-resourced language; we use data from
LibriSpeech \cite{panayotov2015librispeech} and Libri-Light \cite{librilight}, which consists of hundreds of hours of read English.
We consider four low-resource languages: Afrikaans, Setswana, isiXhosa and Sepedi.
Data for the first three are obtained from~\cite{high_qual_tts_2017} (roughly 3.2 h per language), while Sepedi data is obtained from~\cite{spcs}~(10.8~h).

\textbf{VC model:}
For the speech encoder $E_\textrm{CPC}$, we use the pretrained \texttt{CPC-big} model from~\cite{nguyen2020zero}.
The $E_R$ and $E_C$ modules use exactly 
the same architecture as in \cite{tacotron2}, except that the convolutional kernel sizes are 9 instead of 5 and $E_R$ has an additional  
512-dimensional linear output layer.
For the HGST layer, we use $l=3$ sublayers with $h=5$ vectors each, outputting a 512-dimensional style vector $\mathbf{s}$.
With the pretrained $E_\textrm{CPC}$ and quantization module $Q$ fixed, the weights of $E_R$, $E_C$, $D$ and HGST are trained on 275k utterances from the LibriSpeech training set and validated on the remaining utterances in the set.
We use Adam optimization~\cite{adam}, gradient norm clip of 1, a maximum learning rate of $6\cdot10^{-4}$ with the scheduler from~\cite{smith2019:onecycle}, and train for 50 epochs with a batch size of 32.
To stabilize training, we further follow \cite{2021gen_modelling_from_audio}: utterances in a batch are randomly cropped to a fixed length;
the base crop length is linearly increased from 0.6~s to 10.2~s over the training iterations, with the actual crop length uniformly sampled from $\pm20\%$ of the current base value.

\textbf{VC inference:}
In all experiments, the reference utterance is randomly selected from the LibriSpeech training set.
For Afrikaans, isiXhosa and Setswana, we only consider reference utterances from female target speakers since these datasets only contain speech from females.
For English and Sepedi, we consider reference utterances from both male and female speakers. 
At inference time, we convert arbitrary length utterances into 7~s chunks, stitching the results together to produce a final output.

\textbf{SpecAugment:}
We use the SpecAugment settings exactly
as in~\cite{park2019specaugment}.
We also consider whether VC and SpecAugment is complementary: in these experiments
we \textit{first} perform VC and \textit{then} further augment the result with SpecAugment. 
We denote this chained augmentation as VC$\rightarrow$SpecAug.

\textbf{Automatic speech recognition:}
For doing ASR in each of our low-resource languages, we fine-tune the pretrained \texttt{XLSR-53} wav2vec 2.0 model \cite{xlsr-wav2vec2} with a character-level CTC loss~\cite{ctc}.
In validation experiments on English, we use the pretrained wav2vec 2.0 English models from~\cite{wav2vec2.0}.
We train the ASR models using
\texttt{fairseq} \cite{ott2019fairseq} with its 10~min fine-tuning configuration
unless otherwise stated.
Since we are interested in demonstrating the potential \textit{change} in ASR performance, we opt to use greedy CTC decoding without an LM to isolate the effect of augmentation on the acoustic model.
In many real low-resource settings, it might also not be possible to obtain a good LM.
However, to confirm the trends for cases where an LM is available, we also include English results with 
{the official}
4-gram LibriSpeech LM \cite{panayotov2015librispeech}.

\section{Experiments}

\subsection{Experiment 1: Validating voice conversion quality} \label{sec:experiment_1}

We first perform validation experiments to confirm that our VC model performs adequately both on English (its training language) and unseen languages. For the latter, we focus here on Afrikaans and Sepedi, treating the other two languages as completely unseen test languages for our final experiment~(Sec.~\ref{sec:experiment_3}).

To assess intrinsic VC quality, we use two tests.
The first checks
whether linguistic content is retained through a re-synthesis experiment~\cite{2021gen_modelling_from_audio}.
Each test utterance from each language is converted to a target speaker as per Sec.~\ref{sec:exp_setup}.
ASR is then performed on both the input and converted utterances.
If the drop in ASR scores after VC is small in terms of word/character error rate (W/CER), this indicates that VC did not significantly reduce
intelligibility.
Table~\ref{tab:res-resynth} shows the results. 
As expected, the VC data is less intelligible -- higher W/CER for the full model --\ compared to the original audio, with a larger drop in scores on the languages that have not been seen in VC training (Afrikaans and Sepedi).
However, the drop is small enough that the output utterance can still be understood.

The second intrinsic test assesses whether an output utterance matches the target speaker. 
A pretrained speaker verification model~\cite{GE2E}\footnote{{{\scriptsize \url{https://github.com/RF5/simple-speaker-embedding} v0.1}}} is used to check whether the speaker embedding associated with the converted speech is closer to that of the reference or source utterance in terms of their cosine similarity.
We report an error rate: the proportion of converted test-set utterances which are incorrectly closer to the source utterance rather than the reference.
Table~\ref{tab:res-spk-sim} shows the results.
We again see a drop in performance from seen to unseen languages for the full model, where the converted Afrikaans and Sepedi data lead to more errors in capturing the reference speaker identity.
Recall, though, that our goal of data augmentation is to ensure that the converted utterance sounds sufficiently different to the input utterance to increase speaker variety in a limited dataset while still being intelligible so as to aid in ASR training.

\begin{table}[t]
    \renewcommand{\arraystretch}{1.1}
    \centering
    \caption{Re-synthesis results in terms of
    ASR performance on original and converted data.
    The English-trained VC system is applied to
    English, Sepedi and Afrikaans evaluation data.
    Lower W/CER (\%) is better (higher intelligibility).
    }
    \eightpt
    \tablecaptionsep
    \label{tab:res-resynth}
    \begin{tabularx}{\linewidth}{@{}LS[table-format=2.1]S[table-format=2.1]S[table-format=2.1]S[table-format=2.1]S[table-format=2.1]S[table-format=2.1]@{}}
    \toprule
     & \multicolumn{2}{c}{{English}} & \multicolumn{2}{c}{{Afrikaans}} &  \multicolumn{2}{c}{{Sepedi}} \\
    \cmidrule(l){2-3} \cmidrule(l){4-5} \cmidrule(l){6-7}
    
    VC Model & {WER} & {CER} & {WER} & {CER} & {WER} & {CER}\\
    \midrule
    \textit{Original data} & 5.7 & 1.9 & 6.3 & 4.3 & 2.1 & 0.9 \\
    \addlinespace
    Full model & 20.6 & 9.6 & 32.5 & 11.0 & 20.4 & 9.5  \\
    Sans HGST & 21.3 & 9.9 & 34.0 & 11.9 & 21.1 & 9.9 \\
    Sans $Q$ & \ubold 7.1 & \ubold 2.6 & \ubold 17.0 & \ubold 4.6 & \ubold 3.7 & \ubold 1.6 \\
    \bottomrule
    \end{tabularx}
\end{table}

\begin{table}[t]
    \renewcommand{\arraystretch}{1.1}
    \centering
    \caption{Speaker similarity error rates (\%). Lower scores indicate that converted utterances are more often closer to the reference speaker than the source speaker (better~VC).
    }
    \tablecaptionsep
    \eightpt
    \label{tab:res-spk-sim}
    \begin{tabularx}{0.8\linewidth}{@{}LS[table-format=2.1]S[table-format=2.1]S[table-format=2.1]@{}}
    \toprule
    VC Model & {English} & {Afrikaans} & {Sepedi} \\
    \midrule
    Full model & 8.7 & 22.0 & 58.3 \\
    Sans HGST & \ubold 2.3 & \ubold 6.9 & \ubold 34.4 \\
    Sans $Q$ & 99.7 & 99.8 & 99.9 \\
    \bottomrule
    \end{tabularx}
\end{table}

The last two rows in Tables~\ref{tab:res-resynth} and~\ref{tab:res-spk-sim} show the effect when we remove either the HGST
or the normalization-quantization block $Q$ from the full model.
Without the HGST bottleneck, speaker conversion performance improves (Table~\ref{tab:res-spk-sim}), but at the cost of intelligibility (Table~\ref{tab:res-resynth}).
This is because the style vector $\mathbf{s}$ captures more granular reference speaker information (better conversion) but also more content information from the reference (less intelligible).
The opposite happens when we remove the bottleneck $Q$ from the content encoder.
Including both
blocks therefore
gives a balance between intelligibility and conversion~quality, and so we use the full model in the remaining experiments.
\footnote{
{Audio samples: \scriptsize \url{https://rf5.github.io/interspeech2022/}}.}

\subsection{Experiment 2: Validating data augmentation} \label{sec:experiment_2}

Before applying VC cross-lingually (Sec.~\ref{sec:experiment_3}), we first want to test what level of augmentation is beneficial on the seen language (English).
We want to specifically know: How much and in which low-resource settings does real vs augmented data help? And how does it compare or synergize with SpecAugment?
To answer these questions, we fine-tune the \texttt{Base} wav2vec 2.0 English ASR models on 10 min, 1 h, and 10 h of real data from Libri-Light's fine-tuning dataset~\cite{librilight}.
In each setting we train several models: we compare VC, SpecAug, and VC$\rightarrow$SpecAug augmentation (with 100\% and 500\% of additional generated data) to an ASR model trained without any augmented data.

The results are given in Table~\ref{tab:res-aug-levels}.
Looking at the 10 min setting without an
LM, we can make a few observations:
First, using more augmented data appears to improve
performance, with the lowest WER (a 5.3\% absolute improvement over no augmentation) achieved with 500\% additional data generated by chaining VC and SpecAugment.
Second, VC augmentation appears to give similar improvements to SpecAugment when applied independently.
Third, applying both VC and SpecAugment improves performance more than either in isolation.
This shows that the two approaches are complementary on English.

Does this improvement from VC hold in all low-resource settings?
No. With more real data (1 or 10 h), adding more VC-augmented data appears to hinder ASR performance.
VC-augmentation gives more speaker variability but it does so at the cost of
intelligibility (Sec.~\ref{sec:experiment_1}); we suspect higher-resourced settings already have sufficient speaker variability, so the additional (potentially less intelligible) generated data contributes little.
When using an LM for decoding (last 3 columns), the benefits
of any augmentation appear greatly reduced. 
In very low-resource settings,
however, an LM is often not available and
VC still gives a small improvement even though an LM is used.

We can now answer our main research question
in part:
using VC for data augmentation does help in very low-resource contexts on a \textit{seen} language.
We now turn to the question of whether this holds for languages \textit{unseen} during VC training.
As a reminder, our goal is to determine whether VC can improve low-resource ASR, not whether it is the best augmentation method.

\begin{table}[tb]
    \renewcommand{\arraystretch}{1.1}
    \centering
    \caption{
    WERs (\%) on LibriSpeech test data for ASR models trained
    with increasing amounts of VC- and SpecAug-augmented data, with and without 4-gram LM~\mbox{decoding.}
    }
    \tablecaptionsep
    \eightpt
    \label{tab:res-aug-levels}
    \begin{tabularx}{\linewidth}{@{}LcS[table-format=2.1]@{\hspace{0.15cm}}S[table-format=2.1]@{\hspace{0.15cm}}S[table-format=2.1]S[table-format=2.1]@{\hspace{0.15cm}}S[table-format=2.1]@{\hspace{0.15cm}}S[table-format=2.1]@{}}

    \toprule
    & & \multicolumn{3}{c}{{No LM}} & \multicolumn{3}{c}{{LM decoded}} \\
    \cmidrule(l){3-5} \cmidrule(l){6-8}
    
    Augmentation & Amount & {10 min} & {1 h} & {10 h} & {10 min} & {1 h} & {10 h} \\
    \midrule
    None & 0\% & 47.7	& \ubold 30.4 &	13.4 & 17.4 &	\ubold 10.6 &	\ubold 7.5  \\
    VC & 100\% & 43.8 &	32.7 &	13.5 & \ubold 17.2 &	11.4 &	7.6 \\
    VC & 500\% & 43.5 &	34.4 &	14.4 & 17.9 &	11.9 &	8.1 \\
    SpecAug & 100\% & 44.3 &	31.8 &	\ubold 13.1 & 18.8 &	11.2 &	7.6 \\
    SpecAug & 500\% & 43.1 &	34.4 &	13.3 & 17.7	& 12.1 &	7.7 \\
    VC$\rightarrow$SpecAug & 100\%  & 42.5 & 31.3 &	13.2 & 18.5 &	11.2 &	7.6 \\
    VC$\rightarrow$SpecAug & 500\%  & \ubold 42.4 &	35.0 &	14.2 & 18.4	& 12.5 &	8.1 \\
    \bottomrule
    \end{tabularx}
\end{table}

\subsection{Experiment 3: Very low-resource settings}
\label{sec:experiment_3}

In this final experiment we use the English-trained VC system to augment data from four unseen low-resource languages: Afrikaans, Setswana, isiXhosa and Sepedi.
For each language we use a 10 min training set with roughly equally sized validation and test sets covering the rest of the data in each respective language.
There is no speaker overlap between any of the sets.
For each language, we fine-tune \texttt{XLSR-53}~\cite{xlsr-wav2vec2} on the original 10~min of data, and compare this to training with 100\% (10~min) of additional VC and VC$\rightarrow$SpecAug {generated data}.

\begin{table}[!t]
    \renewcommand{\arraystretch}{1.1}
    \centering
    \caption{ASR results (\%) on test data of four low-resource languages when trained on 10 min of real audio data and different amounts of additional VC- and combined VC-SpecAug augmented data. Sepedi$^*$ uses a non-default training procedure.
    }
    \tablecaptionsep
    \eightpt
    \label{tab:res-zs}
    \begin{tabularx}{0.8\linewidth}{@{}lLccc@{}}
    
    \toprule
    Language & Augmentation & Amount & WER & CER \\
    \midrule
    \multirow{3}{*}{Afrikaans}
    & None                    & 0\% & 52.3 & 15.9 \\
    & VC                      & 100\% & \textbf{48.9} & \textbf{15.0} \\ 
    & VC$\rightarrow$SpecAug  & 100\% & 53.5 & 16.5 \\
    \midrule
    \multirow{3}{*}{Setswana} 
    & None                    & 0\% & 68.9 & 26.1 \\
    & VC                      & 100\% & \textbf{65.9} & \textbf{25.1} \\ 
    & VC$\rightarrow$SpecAug  & 100\% & 69.3 & 26.8 \\
    \midrule
    \multirow{3}{*}{isiXhosa}
    & None                    & 0\% & 63.2 & 15.5 \\
    & VC                      & 100\% & \textbf{56.5} & \textbf{13.8} \\ 
    & VC$\rightarrow$SpecAug  & 100\% & 69.3 & 26.8 \\
    \midrule
    \multirow{3}{*}{Sepedi*}
    & None                    & 0\% & 92.6 & 50.7 \\
    & VC                      & 100\% & \textbf{52.8} & \textbf{19.9} \\ 
    & VC$\rightarrow$SpecAug  & 100\% & 97.8 & 69.1 \\
    \bottomrule
    \end{tabularx}
\end{table}

For all four unseen languages, Table~\ref{tab:res-zs} shows
an improvement in ASR performance when 10~min of additional data is generated using the English-trained VC system. 
For isiXhosa, adding the augmented data gives a 7\% absolute improvement in WER.
Chaining VC with SpecAugment, on the other hand,
worsens performance. 
Additional experiments were done with 500\%
of additional VC and VC$\rightarrow$SpecAug data to see if trends followed similar to Sec.~\ref{sec:experiment_2}. However W/CER got worse in all cases compared to training with just the original data.
We also applied SpecAugment in isolation on Afrikaans, which again gave worse performance compared to doing no augmentation.

Some of {Table~\ref{tab:res-zs}'s results} therefore contradicts the results on English in Sec.~\ref{sec:experiment_2}: in Table~\ref{tab:res-aug-levels} SpecAugment and VC gave decent gains when applied in isolation 
and even more so
when chained, while here on unseen languages SpecAugment hampers performance in all cases.
This requires further investigation. Nevertheless, we see on all four low-resource languages that cross-lingual VC augmentation improves ASR performance.

It is worth briefly commenting on the results for Sepedi in Table~\ref{tab:res-zs},
where performance is very poor without augmentation.
We used the
same 10~min training configuration for all languages, but on Sepedi, this configuration failed (WER $>$ 98\%) regardless of whether data augmentation was used or not.
We therefore did validation experiments and made minimal changes to the default configuration, only for Sepedi: we use 1k warmup updates and train for a total of 4k updates.
Using this setting
enabled the ASR model to learn, achieving the test performance in Table~\ref{tab:res-zs}.

Taken together, the results in Table~\ref{tab:res-zs} confirm our hypothesis:
a VC system trained on one well-resourced language (English) can be applied cross-lingually to generate additional data for an unseen low-resource language, improving ASR performance. 
This seem to conflict with~\cite{2019failed_vc_for_asr}, where a similar approach was used but the VC and ASR models were both trained and evaluated on English (Sec.~\ref{sec:related}).
Our study is different in that we do cross-lingual VC, 
and 
also use much less data to fine-tune the ASR models -- since our focus is on very low-resource ASR.

\section{Conclusion}

We have shown that voice conversion (VC) can be used 
for data augmentation
to improve ASR performance in very low-resource settings.
We specifically showed that a VC system trained on one well-resourced language can be used to generate additional training data for unseen low-resource languages.
When labelled resources are very limited (roughly 10~min), 
ASR performance improved on all four low-resource languages considered.
Additional experiments also demonstrated SpecAugment's inability to improve ASR performance in these very low-resource settings. 

Our goal was to show that ASR improvements are possible with VC-based augmentation. 
We did not focus exhaustively on how different design decisions within the VC model itself impacts downstream ASR performance, and the interplay between
VC and traditional augmentation methods remains largely unexplored --\ important directions
for future work.

\section{Acknowledgements}

This work is supported in part by the National Research Foundation of South Africa (grant no. 120409).
Experiments were performed on Stellenbosch University’s HPC Cluster.

\vfill\pagebreak
\bibliographystyle{IEEEtran}

\bibliography{mybib}

\begin{thebibliography}{10}
\providecommand{\url}[1]{#1}
\csname url@samestyle\endcsname
\providecommand{\newblock}{\relax}
\providecommand{\bibinfo}[2]{#2}
\providecommand{\BIBentrySTDinterwordspacing}{\spaceskip=0pt\relax}
\providecommand{\BIBentryALTinterwordstretchfactor}{4}
\providecommand{\BIBentryALTinterwordspacing}{\spaceskip=\fontdimen2\font plus
\BIBentryALTinterwordstretchfactor\fontdimen3\font minus
  \fontdimen4\font\relax}
\providecommand{\BIBforeignlanguage}[2]{{%
\expandafter\ifx\csname l@#1\endcsname\relax
\typeout{** WARNING: IEEEtran.bst: No hyphenation pattern has been}%
\typeout{** loaded for the language `#1'. Using the pattern for}%
\typeout{** the default language instead.}%
\else
\language=\csname l@#1\endcsname
\fi
#2}}
\providecommand{\BIBdecl}{\relax}
\BIBdecl

\bibitem{wav2vec2.0}
A.~Baevski, Y.~Zhou, A.~Mohamed, and M.~Auli, ``wav2vec 2.0: A framework for
  self-supervised learning of speech representations,'' in \emph{NeurIPS},
  2020.

\bibitem{asrsota-zhang2020pushing}
Y.~Zhang, J.~Qin, D.~S. Park, W.~Han, C.-C. Chiu \emph{et~al.}, ``Pushing the
  limits of semi-supervised learning for automatic speech recognition,''
  \emph{arXiv preprint arXiv:2010.10504}, 2020.

\bibitem{hubert2021}
W.-N. Hsu, B.~Bolte, Y.-H.~H. Tsai, K.~Lakhotia, R.~Salakhutdinov
  \emph{et~al.}, ``{HuBERT}: Self-supervised speech representation learning by
  masked prediction of hidden units,'' \emph{arXiv preprint arXiv:2106.07447},
  2021.

\bibitem{besacier2014asr_survey}
L.~Besacier, E.~Barnard, A.~Karpov, and T.~Schultz, ``Automatic speech
  recognition for under-resourced languages: A survey,'' \emph{Speech Commun.},
  vol.~56, pp. 85--100, 2014.

\bibitem{xlsr-wav2vec2}
A.~Conneau, A.~Baevski, R.~Collobert, A.~Mohamed, and M.~Auli, ``Unsupervised
  cross-lingual representation learning for speech recognition,'' \emph{arXiv
  preprint arXiv:2006.13979}, 2020.

\bibitem{Hunt1989data_aug}
M.~Hunt and C.~Lefebvre, ``{A comparison of several acoustic representations
  for speech recognition with degraded and undegraded speech},'' in
  \emph{ICASSP}, 1989.

\bibitem{Lathoud2005data_aug}
G.~Lathoud, M.~Magimai-Doss, B.~Mesot, and H.~Bourlard, ``{Unsupervised
  spectral subtraction for noise-robust ASR},'' in \emph{ASRU}, 2005.

\bibitem{park2019specaugment}
D.~S. Park, W.~Chan, Y.~Zhang, C.-C. Chiu, B.~Zoph \emph{et~al.},
  ``{SpecAugment}: A simple data augmentation method for automatic speech
  recognition,'' in \emph{Interspeech}, 2019.

\bibitem{Mohammadi2017vc_def}
S.~H. Mohammadi and A.~Kain, ``An overview of voice conversion systems,''
  \emph{Speech Commun.}, vol.~88, pp. 65--82, 2017.

\bibitem{hifi-gan}
J.~Kong, J.~Kim, and J.~Bae, ``{HiFi-GAN}: Generative adversarial networks for
  efficient and high fidelity speech synthesis,'' in \emph{NeurIPS}, 2020.

\bibitem{vcc2020}
Y.~{Zhao}, W.-C. {Huang}, X.~{Tian}, J.~{Yamagishi}, R.~K. {Das} \emph{et~al.},
  ``Voice conversion challenge 2020: Intra-lingual semi-parallel and
  cross-lingual voice conversion,'' \emph{arXiv preprint arXiv:2006.13979},
  2020.

\bibitem{triple_bottle_discrete_disentangle}
A.~Polyak, Y.~Adi, J.~Copet, E.~Kharitonov, K.~Lakhotia \emph{et~al.}, ``Speech
  resynthesis from discrete disentangled self-supervised representations,'' in
  \emph{Interspeech}, 2021.

\bibitem{autovc}
K.~Qian, Y.~Zhang, S.~Chang, X.~Yang, and M.~Hasegawa-Johnson, ``{AutoVC}:
  Zero-shot voice style transfer with only autoencoder loss,'' in \emph{PMLR},
  2019.

\bibitem{stargan-zsvc}
M.~Baas and H.~Kamper, ``{StarGAN-ZSVC}: Towards zero-shot voice conversion in
  low-resource contexts,'' in \emph{SACAIR}, 2020.

\bibitem{hst}
X.~An, Y.~Wang, S.~Yang, Z.~Ma, and L.~Xie, ``Learning hierarchical
  representations for expressive speaking style in end-to-end speech
  synthesis,'' in \emph{ASRU}, 2019.

\bibitem{yang2021cross}
Z.~Yang, W.~Zhang, Y.~Liu, and X.~Xing, ``Cross-lingual voice conversion with
  disentangled universal linguistic representations,'' in \emph{Interspeech},
  2021.

\bibitem{jaitly2013vltp_aug}
N.~Jaitly and G.~E. Hinton, ``Vocal tract length perturbation {(VTLP)} improves
  speech recognition,'' in \emph{ICML}, 2013.

\bibitem{tts_data_aug}
A.~Laptev, R.~Korostik, A.~Svischev, A.~Andrusenko, I.~Medennikov
  \emph{et~al.}, ``You do not need more data: Improving end-to-end speech
  recognition by text-to-speech data augmentation,'' in \emph{CISP-BMEI}, 2020.

\bibitem{how_low_is_low_resource}
M.~A. Hedderich, L.~Lange, H.~Adel, J.~Str{\"o}tgen, and D.~Klakow, ``A survey
  on recent approaches for natural language processing in low-resource
  scenarios,'' in \emph{NAACL}, 2021.

\bibitem{Bellegarda1994spk_norm_vc}
J.~Bellegarda, P.~de~Souza, A.~Nadas, D.~Nahamoo, M.~Picheny \emph{et~al.},
  ``The metamorphic algorithm: a speaker mapping approach to data
  augmentation,'' \emph{IEEE/ACM TASLP}, vol.~2, no.~3, pp. 413--420, 1994.

\bibitem{2015_simple_vc_for_asr}
X.~Cui, V.~Goel, and B.~Kingsbury, ``Data augmentation for deep neural network
  acoustic modeling,'' \emph{IEEE/ACM TASLP}, vol.~23, no.~9, pp. 1469--1477,
  2015.

\bibitem{2021vc_for_tts}
G.~Huybrechts, T.~Merritt, G.~Comini, B.~Perz, R.~Shah \emph{et~al.},
  ``Low-resource expressive text-to-speech using data augmentation,'' in
  \emph{ICASSP}, 2021.

\bibitem{2020children_vc_asr}
S.~Shahnawazuddin, N.~Adiga, K.~Kumar, A.~Poddar, and W.~Ahmad, ``Voice
  conversion based data augmentation to improve children’s speech recognition
  in limited data scenario,'' in \emph{Interspeech}, 2020.

\bibitem{crosslingual_vc}
S.~Zhao, T.~H. Nguyen, H.~Wang, and B.~Ma, ``Towards natural bilingual and
  code-switched speech synthesis based on mix of monolingual recordings and
  cross-lingual voice conversion,'' in \emph{Interspeech}, 2020.

\bibitem{2019failed_vc_for_asr}
G.~Keskin, T.~Lee, C.~Stephenson, and O.~H. Elibol, ``Measuring the
  effectiveness of voice conversion on speaker identification and automatic
  speech recognition systems,'' \emph{arXiv preprint arXiv:1905.12531}, 2019.

\bibitem{nguyen2020zero}
T.~A. Nguyen, M.~de~Seyssel, P.~Rozé, M.~Rivière, E.~Kharitonov
  \emph{et~al.}, ``The zero resource speech benchmark 2021: Metrics and
  baselines for unsupervised spoken language modeling,'' \emph{arXiv preprint
  arXiv:2011.11588}, 2020.

\bibitem{niekerk21_interspeech}
B.~van Niekerk, L.~Nortje, M.~Baas, and H.~Kamper, ``Analyzing speaker
  information in self-supervised models to improve zero-resource speech
  processing,'' in \emph{Interspeech}, 2021.

\bibitem{lstm}
S.~Hochreiter and J.~Schmidhuber, ``{Long Short-term Memory},'' \emph{Neural
  Comput.}, vol.~9, pp. 1735--80, 1997.

\bibitem{tacotron2}
J.~Shen, R.~Pang, R.~J. Weiss, M.~Schuster, N.~Jaitly \emph{et~al.}, ``Natural
  {TTS} synthesis by conditioning wavenet on mel spectrogram predictions,'' in
  \emph{ICASSP}, 2018.

\bibitem{panayotov2015librispeech}
V.~Panayotov, G.~Chen, D.~Povey, and S.~Khudanpur, ``Librispeech: an {ASR}
  corpus based on public domain audio books,'' in \emph{ICASSP}, 2015.

\bibitem{librilight}
J.~{Kahn}, M.~{Rivière}, W.~{Zheng}, E.~{Kharitonov}, Q.~{Xu} \emph{et~al.},
  ``{Libri-Light}: A benchmark for {ASR} with limited or no supervision,'' in
  \emph{ICASSP}, 2020.

\bibitem{high_qual_tts_2017}
D.~van Niekerk, C.~van Heerden, M.~Davel, N.~Kleynhans, O.~Kjartansson
  \emph{et~al.}, ``Rapid development of {TTS} corpora for four {South African}
  languages,'' in \emph{Interspeech}, 2017.

\bibitem{spcs}
T.~I. Modipa, M.~H. Davel, and F.~de~Wet, ``Implications of {Sepedi/English}
  code switching for {ASR} systems,'' in \emph{PRASA}, 2013.

\bibitem{adam}
D.~P. Kingma and J.~Ba, ``Adam: A method for stochastic optimization,'' in
  \emph{ICLR}, 2015.

\bibitem{smith2019:onecycle}
L.~N. Smith and N.~Topin, ``Super-convergence: Very fast training of neural
  networks using large learning rates,'' in \emph{SPIE AIML MDO}, 2019.

\bibitem{2021gen_modelling_from_audio}
K.~Lakhotia, E.~Kharitonov, W.-N. Hsu, Y.~Adi, A.~Polyak \emph{et~al.},
  ``Generative spoken language modeling from raw audio,'' \emph{arXiv preprint
  arXiv:2102.01192}, 2021.

\bibitem{ctc}
A.~Graves, S.~Fern\'{a}ndez, F.~Gomez, and J.~Schmidhuber, ``Connectionist
  temporal classification: Labelling unsegmented sequence data with recurrent
  neural networks,'' in \emph{ICML}, 2006.

\bibitem{ott2019fairseq}
M.~Ott, S.~Edunov, A.~Baevski, A.~Fan, S.~Gross \emph{et~al.}, ``fairseq: A
  fast, extensible toolkit for sequence modeling,'' in \emph{NAACL-HLT}, 2019.

\bibitem{GE2E}
L.~Wan, Q.~Wang, A.~Papir, and I.~L. Moreno, ``Generalized end-to-end loss for
  speaker verification,'' in \emph{ICASSP}, 2018.

\end{thebibliography}

\end{document}